\documentclass[traditabstract]{aa}
\bibpunct{(}{)}{;}{a}{}{,} 
\usepackage{graphicx}
\usepackage[varg]{txfonts}
\begin{document}

\title{Constraining the geometry and kinematics of the quasar broad
  emission line region using gravitational microlensing.\\
  II. Comparing models with observations in the lensed
  quasar HE0435-1223}
\author{D. Hutsem\'ekers\inst{1,}\thanks{Senior Research Associate F.R.S.-FNRS},
        L. Braibant\inst{1,},
        D. Sluse\inst{1},
        R. Goosmann\inst{2}
        }
\institute{
    Institut d'Astrophysique et de G\'eophysique,
    Universit\'e de Li\`ege, All\'ee du 6 Ao\^ut 19c, B5c,
    4000 Li\`ege, Belgium
    \and 
    Observatoire Astronomique de Strasbourg,
    Universit\'e de Strasbourg, Rue de l’Universit\'e 11,
    F-67000 Strasbourg, France
    }
\date{Received ; accepted: }
\titlerunning{Microlensing of the quasar broad line region. II. HE0435-1223} 
\authorrunning{D. Hutsem\'ekers et al.}
\abstract{The quadruply lensed quasar HE0435-1223 shows a clear microlensing effect that affects differently the blue and red wings of the H$\alpha$ line profile in its image D. To interpret these observations, and constrain the broad emission line region (BLR) properties, the effect of gravitational microlensing on quasar broad emission line profiles and their underlying continuum has been simulated considering representative BLR models and microlensing magnification maps. The amplification and distortion of the H$\alpha$ line profile, characterized by a set of four indices, can be reproduced by the simulations.  Although the constraints on the BLR models set by the observed single-epoch microlensing signal are not very robust, we found that flattened geometries (Keplerian disk and equatorial wind) can more easily reproduce the observed line profile deformations than a biconical polar wind. With an additional independent constraint on the size of the continuum source, the Keplerian disk model of the H$\alpha$ BLR is slightly favored.
}
\maketitle
%
%
%
\section{Introduction}
\label{sec:intro}

Recent studies have shown that line profile distortions are commonly observed in gravitationally lensed quasar spectra \citep{2004Richards, 2005Wayth, 2007Sluse, 2011Sluse, 2012Sluse, 2011ODowd, 2013Guerras, 2014Braibant, 2016Braibant,2017Motta}. They are usually attributed to microlensing, that is, differential magnification of spatially and kinematically separated subregions of the broad emission line region (BLR). Line profile distortions can thus provide information on the quasar BLR structure. These line profile deformations are most often detected as red/blue or wings/core distortions \citep{2012Sluse}.

In a previous paper (\citealt{2017Braibant}, hereafter Paper~I; see also \citealt{2018Braibant}), we computed the effect of gravitational microlensing on quasar broad emission line profiles and their underlying continuum, considering several representative BLR models and generic microlensing magnification maps. To analyze the large amount of simulated line profiles, the effects of microlensing have been quantified using four observables:  $\mu^{cont}$, the magnification of the continuum, $\mu^{BLR}$, the total magnification of the broad emission line, as well as two indices sensitive to red/blue and wings/core line profile distortions, $RBI$ and $WCI$, respectively.  In particular, we built $(WCI,RBI)$ diagrams that can serve as ``diagnostic diagrams'' to discriminate the different BLR models on the basis of quantitative measurements.

Using infrared spectra of the quadruply lensed quasar HE0435-1223 acquired in 2009 with the spectrograph SINFONI at the ESO Very Large Telescope, \citet{2014Braibant}  detected in image D a clear microlensing effect that affects the blue and red wings of the H$\alpha$ line profile differently. This observation suggests that the H$\alpha$ line originates from a rotating disk.

To substantiate this result, we now compare these observations with simulations. We first characterize  the microlensing effect in the  H$\alpha$ line profile of HE0435-1223 by measuring the four indices $\mu^{cont}$, $\mu^{BLR}$, $RBI$, and $WCI$ defined in Paper~I (Sect.~\ref{sec:he0435_mu_rbi_wci}). These indices are then compared to those obtained from the simulations with the aim to constrain the BLR properties (Sect.~\ref{sec:he0435_comparison_with_modeling}). Conclusions form the last section.

\section{Measurement of microlensing amplification and distortion indices}
\label{sec:he0435_mu_rbi_wci}

\begin{figure}
\centering
\resizebox{\hsize}{!}{\includegraphics*{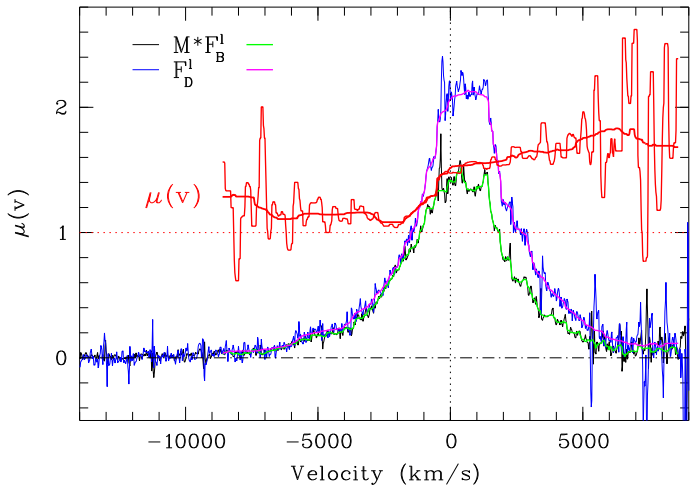}}
\caption{Flux density ratio  $\mu(v) = F^l_{\text D} / (M \times F^l_{\text B})$ computed for the H$\alpha$ emission line observed in the lensed quasar HE0435-1223. This ratio is illustrated as a function of the Doppler velocity  over the useful $[-8600,8600]$ km~s$^{-1}$ velocity range (red lines). The continuum-subtracted spectra of images B and D ($M \times F^l_{\text B}$ and $F^l_{\text D}$, respectively) are superimposed, on an arbitrary flux scale (thin black and blue lines). These spectra were corrected for instrumental spikes and smoothed using a ${220}$~km~s$^{-1}$ -wide median filter in the velocity range of interest (thick green and magenta lines).  $\mu(v)$ is shown with the ${220}$~km~s$^{-1}$ smoothing and a stronger one using a ${1500}$~km~s$^{-1}$-wide median filter.}
\label{fig:muwave}
\end{figure}

\citet{2014Braibant} have shown that microlensing in HE0435-1223\footnote{The source and lens redshifts are $z_S$ = 1.693 and $z_L$ = 0.454, respectively \citep{2012Sluse}. To compute the lensing parameters, we adopted a flat $\Lambda$CDM cosmology with $H_0 = 68$ km~s$^{-1}$ Mpc$^{-1}$ and $\Omega_m$ = 0.31.} modifies the spectrum of image D, leaving image B unaffected. We thus consider the spectrum of image D as the microlensed spectrum, and the spectrum of image B as the non-microlensed reference spectrum. The magnification of the continuum emission underlying the H$\alpha$ line was estimated to be $\mu^{cont}=1.68 \pm 0.10$.

The total microlensing-induced magnification of the emission line, $\mu^{BLR}$, is obtained as the ratio between the H$\alpha$ line flux in image D and the  H$\alpha$ line flux in image B, corrected for the D/B macro-amplification ratio $M = 0.47 \pm 0.03$ determined in \citet{2014Braibant} from the same data set:
\begin{equation}
\mu^{BLR} = \frac{1}{M} \frac{ \int_{v_{-}}^{v_{+}} \, F^l_{\text{D}} \, (v) \, dv}{\int_{v_{-}}^{v_{+}} F^l_{\text{B}} \, (v) \, dv}  \;  .
\label{eq:cmp-mublr}
\end{equation}
The Doppler velocities $v$ are computed using the redshift $z_S$ = 1.693 determined from the \ion{Mg}{ii} emission line \citep{2012Sluse}. Flat continua are subtracted from the spectra of images B and D, individually, in order to extract the line flux densities $F^l_{\text B}(v)$ and $F^l_{\text D}(v)$. In each image spectrum, the continuum is assessed as the median value of the flux density in the range $[-18000,-14500]$ km~s$^{-1}$. The line flux is then integrated over the whole line profile, that is, over the $[v_{-},v_{+}]$ = $[-8600,8600]$ km~s$^{-1}$ velocity range. The total microlensing magnification of the BLR is $\mu^{BLR}= 1.30 \pm 0.17$. 

The red/blue $RBI$ and wings/core $WCI$ indices characterizing the line profile distortions are estimated using
\begin{equation}
RBI = \frac{\int_0^{v_{+}} \log \left( \mu(v) \right) dv}{\int_0^{v_{+}} dv} - \frac{\int_{v_{-}}^0 \log \left( \mu(v) \right) dv}{\int_{v_{-}}^0 dv} \; 
\label{eq:bluered_param}
\end{equation}
and
\begin{equation}
WCI = \frac{\int_{v_{-}}^{v_{+}} \mu(v)/ \mu(v=0) \, dv}{\int_{v_{-}}^{v_{+}} dv} \; ,
\label{eq:wingcore_param}
\end{equation}
where
\begin{equation}
\mu \, (v) =  \frac{1}{M}  \frac{F^l_{\text{D}} \, (v) }{F^l_{\text{B}} \, (v)} \; .
\label{eq:muv}
\end{equation}
RBI is sensitive to the symmetry of the deformations of the line profile. It takes non-null values when the effect of microlensing on the blue and red parts of the line is asymmetric. A positive $RBI$ indicates that the red part of the line profile undergoes, on average, a larger magnification than the blue one.  WCI indicates whether the whole emission line is, on average, more or less affected by microlensing than its center. In particular, a significantly higher magnification of the line wings with respect to the line core results in values of $WCI$ larger than one. As they are defined, both $RBI$ and $WCI$ are independent of $M$.

In Fig.~\ref{fig:muwave}, $\mu(v)$ is plotted as red lines with two different smoothings. With respect to the line core value, $\mu(v)$ drops in the blue side of the H$\alpha$ line and slowly increases on the red side, indicating that the blue part of the emission line is less affected by microlensing than the red part.  Using Eqs.~\ref{eq:bluered_param}-\ref{eq:muv}, we measure $RBI= 0.15 \pm 0.02$ and $WCI=1.09 \pm 0.17$. The errors on $\mu^{BLR}$, $RBI$,  and $WCI$ are obtained by propagating the uncertainty on the line flux densities. That uncertainty is computed as the quadratic sum of the error on the total (line + continuum) flux density and the error on the continuum estimate, the latter being taken as the standard deviation of the continuum values in the  $[-18000,-14500]$ km~s$^{-1}$ velocity range. We note that a reasonable smoothing of the spectra when calculating $RBI$ and $WCI$ does not change the results, as expected for integrated quantities.  In Fig.~\ref{fig:muwave} $\mu(v)$ is smoothed in two different ways, and the corresponding values of $RBI$ and $WCI$ do not change by more than 5\%, which is much lower than the quoted error bars.

The significantly positive value of $RBI$ indicates that microlensing magnifies the receding velocities to a larger extent than the approaching ones. On the other hand, $WCI$ that is equal to one within the uncertainties indicates that the microlensing effect does not induce significant wings/core distortions.

\begin{figure*}
\centering
\resizebox{\hsize}{!}{%
\includegraphics*[trim={0 0  5 20mm},clip]{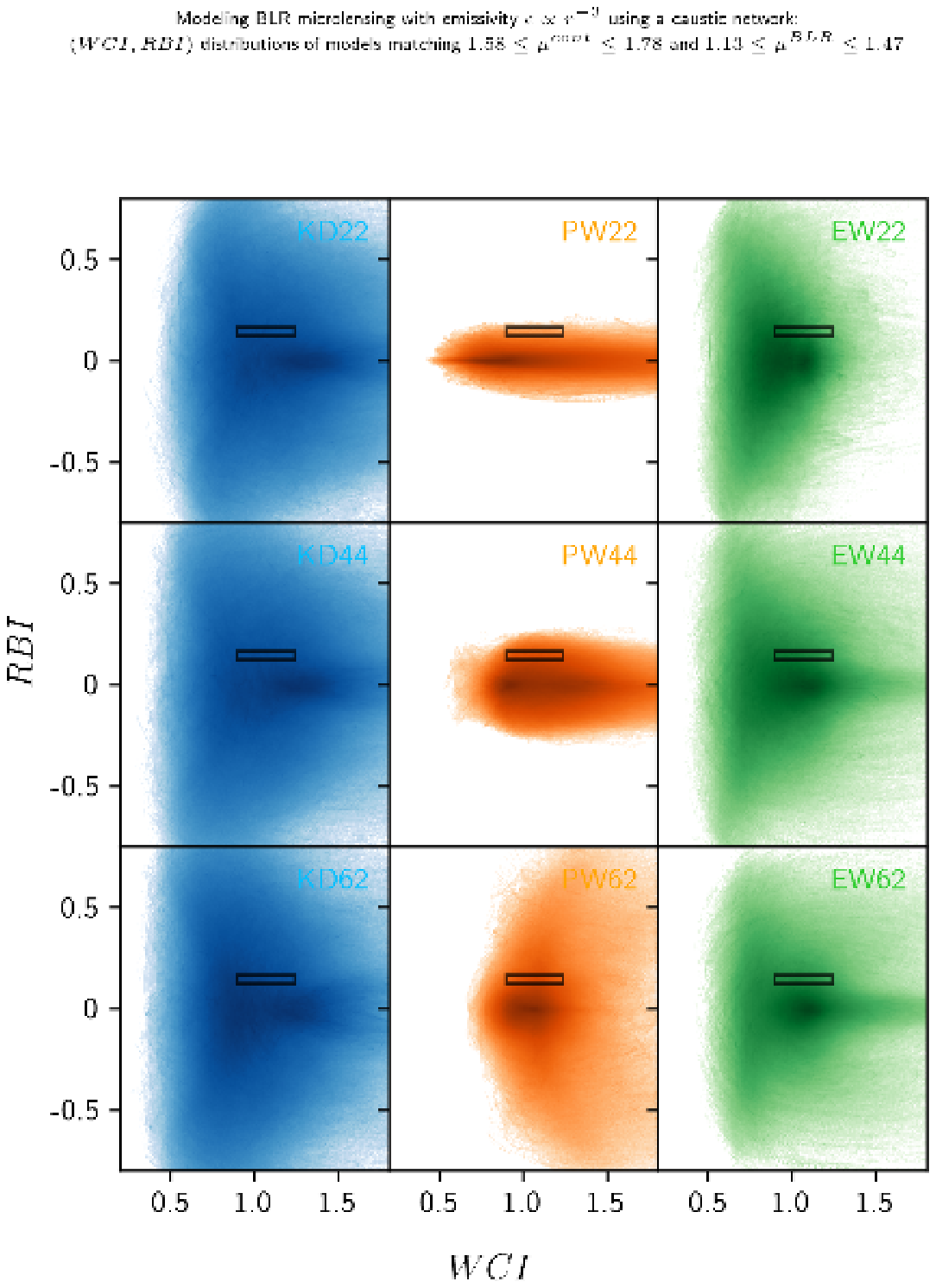}%
\includegraphics*[trim={0 0 10 20mm},clip]{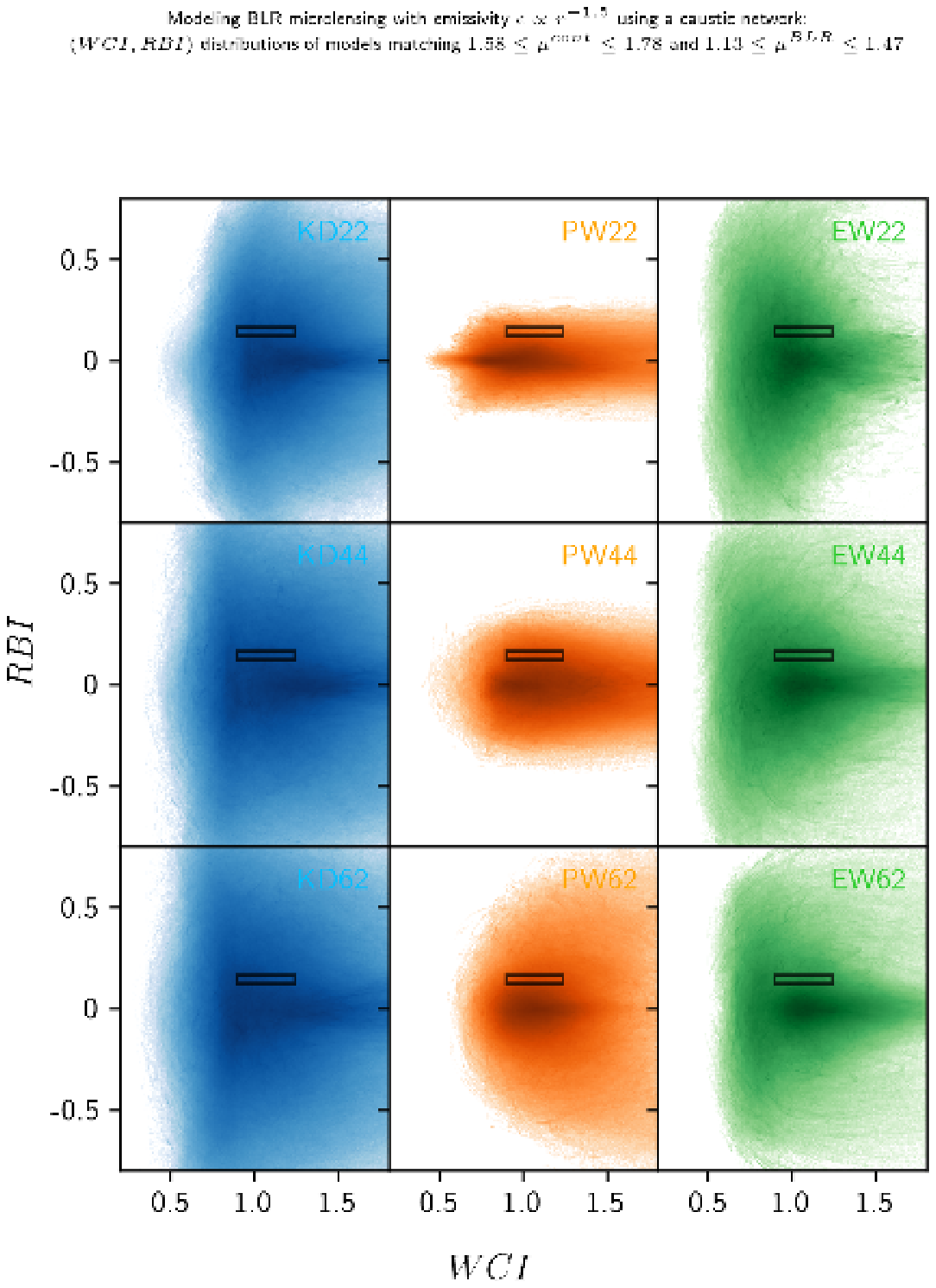}}
\caption{Two-dimensional histograms of simulated $(WCI,RBI)$. These indices were  measured from simulated line profiles that arise from the BLR models KD, PW, and EW seen at inclinations 22\degr, 44\degr, and 62\degr. The BLR models that have an emissivity $\epsilon_0 \, (r_{\text{in}}/r)^q$ that sharply decreases with radius, i.e., $q=3$, are illustrated in the left panel, while those characterized by a slowly decreasing emissivity, i.e., $q=1.5$, are illustrated in the right panel. Each model also contains a continuum-emitting disk seen under the same inclination. All continuum source radii and BLR sizes used in the simulations are considered but only the models that reproduce the measured $\mu^{cont}$ and $\mu^{BLR}$ values are shown, i.e.,  $1.58 \leq \mu^{cont} \leq 1.78$ and $1.13 \leq \mu^{BLR} \leq 1.47$. The color map is logarithmic. The measured $WCI$ and $RBI$ values with their uncertainties are plotted as a small grey rectangle superimposed on the simulated $(WCI,RBI)$ distributions.}
\label{fig:he0435_mod2}
\end{figure*}

\section{Comparison with simulations}
\label{sec:he0435_comparison_with_modeling}

Following Paper~I, we  investigate the effect of gravitational microlensing on quasar broad emission line profiles and their underlying continuum, by convolving in the source plane the emission from representative BLR models with microlensing magnification maps. For the BLR, we consider rotating Keplerian disks (KD), as well as polar (PW) and equatorial (EW) wind models that are radially accelerated. The models have inclinations with respect to the line of sight $i$ = 22\degr, 34\degr, 44\degr, 62\degr.  We investigate nine inner radius values for each BLR model: $r_{\text{in}}=$ 0.1, 0.125, 0.15, 0.175, 0.2, 0.25, 0.35, 0.5, and 0.75 $r_E$, where $r_E$ is the Einstein radius. The outer radius is fixed to $r_{\text{out}} = 10 \, r_{\text{in}}$. The BLR models are assumed to have an emissivity $\epsilon = \epsilon_0 \, (r_{\text{in}}/r)^q$ that either sharply decreases with radius, with $q=3$, or more slowly, with $q=1.5$. Each model also contains a continuum-emitting uniform disk seen under the same inclination as the BLR, and with an outer radius fixed at nine different values: $r_s =$ 0.1, 0.15, 0.2, 0.25, 0.3, 0.4, 0.5, 0.6, and 0.7 $r_E$. For further details, we refer to Paper~I.

\subsection{Simulations with a caustic network}

Modeling the effect of microlensing on the BLR was achieved using a caustic network specific to image D of HE0435-1223. This complex magnifying structure was computed using the \texttt{microlens} code \citep{1999Wambsganss}, and considering a convergence of $\kappa_s = $ 0.124 for matter in compact objects and $\kappa_c = $ 0.466 for continuously distributed matter, with an external shear $\gamma =$ 0.640. These values were taken from the macro-model parameters at the location of image~D given in \citet{2012Sluse}, the fraction of compact matter being estimated by \citet{2015Jimenez}. The magnification map extends over a $200 \, r_E \times 200 \, r_E$ area of the source plane and was sampled by $20000 \times 20000$ pixels. To limit the impact of preferential alignment between the symmetry axes of the BLR models and the caustic network, the map was rotated by 0\degr, 30\degr, 45\degr, 60\degr\ and 90\degr , and only the central $10000 \times 10000$ pixel parts of the rotated maps were used.

Distorted line profiles are obtained from the convolution of the monochromatic images of the synthetic BLR by the caustic network map, and computed for each position of the BLR on the magnification maps, generating $\sim 10^8$ simulated profiles per map and BLR model combination. Given the huge amount of simulated line profiles, we focus our analysis on the indices $\mu^{cont}$, $\mu^{BLR}$, $RBI$, and $WCI$ that characterize the continuum and line profile magnification and distortions, and that can be directly compared to the observations. As shown in Appendix~A, simulated magnification profiles verifying the observational constraints on $WCI$ and $RBI$ reproduce the observed $\mu (v)$  profile reasonably well.

Figure~\ref{fig:he0435_mod2} displays the $(WCI,RBI)$ distributions of the simulated microlensed line profiles that arise from the magnification of the different BLR models by the caustic network. Only simulations that match the $\mu^{cont}$ and $\mu^{BLR}$ values measured in image D are represented. The BLR models are grouped according to their geometry, inclination and emissivity. Models computed with the inclination $i = 34\degr$ are not illustrated since they produce $(WCI,RBI)$ distributions that are intermediate between  $i = 22\degr$ and  $i = 44\degr$, whatever the  geometry/kinematics. The ranges of $WCI$ and $RBI$ values that correspond to the observed distortion of the H$\alpha$ line profile are indicated by a small rectangle superimposed on top of the simulated $(WCI,RBI)$ distributions.

We immediately see that most models generate a wide range of distortions (i.e., $WCI$ and $RBI$ values), and that flattened geometries, that is, KD and EW, can reproduce the observations much more easily whatever the inclination, in particular the high $RBI$ value that characterizes the asymmetric distortion of H$\alpha$. The biconical PW reproduces the observed distortions more marginally, especially at low inclinations. No significant difference can be observed between simulations performed with the $q=3$ and $q=1.5$ emissivity indices. 

\subsection{Counts and probabilities}

In order to quantify the differences between the BLR models reproducing the measurements, we need to calculate the posterior probability distribution associated with each model G (where G = KD, PW, or EW) given the data $\bf{d_{obs}}$. In the present case, the data are summarized by four observable quantities: $\mu^{cont}$, $\mu^{BLR}$, $RBI$, and $WCI$. The posterior probability for a model G is expressed as
\begin{eqnarray}
P(G, \boldsymbol{\theta}, \boldsymbol{\eta}\,|\, {\bf d} ) \propto L( {\bf d} \,| \, \boldsymbol{\theta}, \boldsymbol{\eta}) \times \pi(\boldsymbol{\theta}) \times \pi(\boldsymbol{\eta}),  
\end{eqnarray}
where we have separated the parameters ${\boldsymbol{\theta}}$, associated with the BLR model $G$, and the microlensing parameters $\boldsymbol{\eta}$. The BLR models considered in this work share the same set of physical parameters ${\boldsymbol{\theta}} = (i, r_{\text{in}}, q, r_s),$ namely the inclination angle $i$, the inner radius of the BLR $r_{\text{in}}$, the emissivity index $q$, and the radius of the continuum source $r_s$. The parameters $\boldsymbol{\eta}$ are the position $(x, y)$ of the source in the micro-magnification map and the orientation of the microlensing map with respect to the symmetry axis of the BLR model.  Finally, $\pi({\boldsymbol{\theta}})$ and $\pi({\boldsymbol{\eta}})$ are the priors on the model parameters. These priors are identical for the different models considered.  

For each set of parameters, we compute the likelihood as
\begin{eqnarray}
L( {\bf{d_{obs}}}\,|\,{\boldsymbol{\theta}}, {\boldsymbol{\eta}}) =  \exp \left( - 0.5 \sum_{j = 1}^{4}  \left( \frac{ d_j-d_{{\rm obs}, j}}{\sigma_{d_{{\rm obs}, j}}} \right) ^{2}    \right) \; ,
\end{eqnarray}
where $d_{{obs}, j}$ is the $j^{th}$ observable measured with an uncertainty $\sigma_{d_{{\rm obs}, j}}$, and $d_j$ is the value predicted by the model. 

Since we cannot accurately estimate $r_{\text{in}}$ and $r_s$ from our single epoch observations, and the emissivity index cannot be readily discriminated from the simulations (as seen in Fig.~\ref{fig:he0435_mod2}), we marginalize the likelihood over those three quantities, as well as over the microlensing parameters $\boldsymbol{\eta}$. The marginalized likelihood for $\theta_k = i$ and a specific BLR model G is computed as
\begin{eqnarray}
L({\bf{d}}\,|\,i, G) =   \sum_{r_{\text{in}}} \sum_{q}  \sum_{r_s} \sum_{n} L({\bf{d_{obs}}} \,|\, r_{\text{in}}, q, r_s, {\boldsymbol{\eta}})  / n \; ,
\end{eqnarray}
where $n$ is the number of different choices of ${\boldsymbol{\eta}}$ (i.e., microlensing parameters), corresponding to a given set of BLR parameters $(r_{\text{in}}, q, r_s)$. This number slightly differs for different $r_{\text{in}}$ because the usable part of the convolved caustic network map decreases as the size of the BLR increases. 

Since the different BLR models share the same parameters and associated priors, we can quantify their relative efficiency in reproducing the data by comparing the likelihoods. More specifically, it is sufficient to normalize the marginalized likelihood by the sum of the likelihoods associated to each model $G$ for each inclination $i$\footnote{An alternative procedure would be to calculate the evidence associated to each model, but due to the coarse sampling of the parameter space this approach is practically unworkable. }. This yields the relative probability
\begin{eqnarray}
\mathcal{P} (G, i \,| \, {\bf{d}}) = L({\bf{d}}\, |\, G, i) / \sum_{G} \sum_{i} L({\bf{d}} | G, i) \; 
\end{eqnarray}
that a given model $(G, i)$ can more easily reproduce the four observables $\mu^{cont}$, $\mu^{BLR}$, $RBI$, and $WCI$ than another model. The values are given in Table~\ref{tab:proba}\footnote{We also considered the probabilities for models computed with the emissivity indices $q=3$ and $q=1.5$ separately, but no significant difference was found.}.

As we can see from Table~\ref{tab:proba}, the constraints set by the microlensing signal do not enable the  unambiguous discrimination of the BLR models. Nervertheless, models with flattened geometries (KD and EW) can more easily reproduce the observed line profile deformations than the PW. The KD and EW models do not show any preferred inclination.

\begin{table}[t]
\caption{Probability (in \%) of the different models}
\label{tab:proba}
\centering
\begin{tabular}{lccccccc}
\hline\hline
   &  \multicolumn{3}{c}{All $r_s$}  &  & \multicolumn{3}{c}{$r_s \geq$ 0.6 $r_E$} \\
\hline
          & KD & PW & EW & & KD & PW & EW  \\
\hline
          22\degr         & 12 &  1 & 11 & & 14 &  0 &  8 \\
          34\degr         & 10 &  3 & 11 & & 13 &  1 &  8 \\
          44\degr         & 10 &  7 &  9 & & 13 &  7 &  7 \\
          62\degr         & 10 &  9 &  8 & & 13 &  8 &  6 \\
          All $i$         & 42 & 20 & 38 & & 54 & 16 & 30 \\
\hline
\end{tabular}
\end{table}

\subsection{Adding a constraint on the continuum source size}

Independent knowledge of the size of the continuum source may constitute an interesting additional constraint when comparing observations to simulations. The size of the continuum source controls the range of $\mu^{cont}$ values and fixes the smallest possible inner radius of the BLR. Increasing the size of the continuum source $r_s$ implies more extended BLRs since $r_{\text{in}} \geq r_s$, thus reducing the range of possible variations of $\mu^{BLR}$, $WCI$, and $RBI$.

Several estimates of the size of the accretion disk in HE0435-1223 have been derived from microlensing studies \citep{2010Morgan,2011Mosquera,2011Blackburne,2014Blackburne,2014Jimenez,2018Fian}. At the wavelength of H$\alpha$, half-light radius estimates range from $R_{1/2} \simeq 6$ to $30$ light days. These values are computed for a microlens mass $M$ = $0.3 M_{\sun}$, and using  $R_{1/2} (\text{H}\alpha) = R_{1/2} (\text{UV}) \times (\lambda_{\text{H}\alpha} / \lambda_{\text{UV}}) ^p$, where $R_{1/2} (\text{UV})$ stands for the measured rest-frame UV half-light radius. When $p$ was not simultaneously determined, we assumed $p=4/3$. Considering that the HE0435-1223 lensed system is characterized by an Einstein radius $r_E \simeq 11.5$ light days ($M$ = $0.3 M_{\sun}$), this is equivalent to a uniformly emitting accretion disk of outer radius $r_s \simeq 1.4 \, R_{1/2} \,$ that ranges from $r_s \simeq 0.7$ to $4 \, r_E$. 

Probabilities obtained when restricting the analysis to $r_s \geq 0.6 \, r_E$ are also reported in Table~\ref{tab:proba}. Even with this restriction, several thousands of configurations can still reproduce the observations. 
This additional constraint slightly reinforces the flattened geometries and in particular the KD as the model that most easily reproduces the observed line profile deformations.

\section{Conclusions}
\label{sec:conclu}

The observed H$\alpha$ line profile amplification and distortion characterized by the four indices $\mu^{cont}$, $\mu^{BLR}$, $WCI$, and $RBI$ are well reproduced by our simulations based on simple BLR and microlensing models. 
We found that flattened geometries (KD and EW) can more easily reproduce the observed line profile deformations than a biconical polar wind. Adding a constraint on the size of the continuum source from independent measurements allows us to restrict the number of possible configurations matching the observations. In that case the KD model is slightly favored. The constraints set by the microlensing signal, observed at a single epoch, cannot discriminate the BLR models more robustly. These results support our previous conclusions based on spectral disentangling \citep{2014Braibant}.

Since models with different geometry and kinematics sample different regions of the caustic pattern, multi-epoch observations of the microlensing effects appear as one of the next steps to further discriminate the BLR models. Alternatively, by simultaneously taking into account the distortions induced by the same magnification pattern on different emission lines, like \ion{C}{iv} and H$\alpha$ in the Einstein Cross \citep{2016Braibant}, stronger constraints on the BLR models can be derived.

\bibliographystyle{aa}
\bibliography{references}

\begin{appendix}
\section{Examples of simulated magnification profiles $\mu(v)$}
\label{sec:appendix}

Given the difficulty of accurately measuring the whole magnification profile $\mu(v)$ and the fact that only large contiguous parts of the line profile are observed to be magnified \citep[see  also][]{2012Sluse}, we use the global indices $\mu^{BLR}$, $WCI$, and $RBI$ to characterize line profile distortions due to microlensing. Using indices that capture the major line profile distortions also helps to handle the large amount of simulated data more easily.

\begin{figure*}[b]
\centering
\resizebox{\hsize}{!}{
\includegraphics*[trim={0 0 0 0mm},clip]{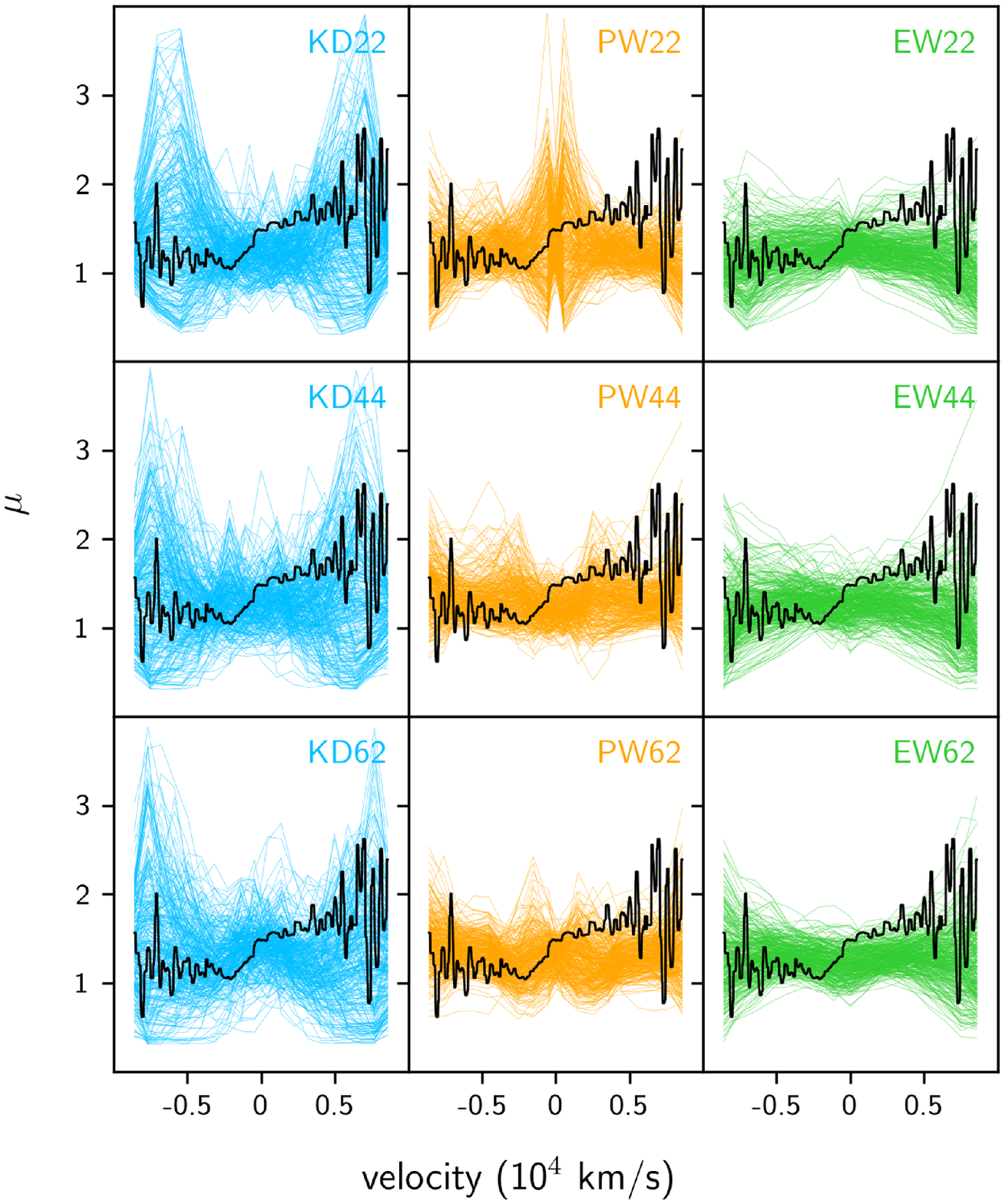}%
\includegraphics*[trim={20 0 0 0mm},clip]{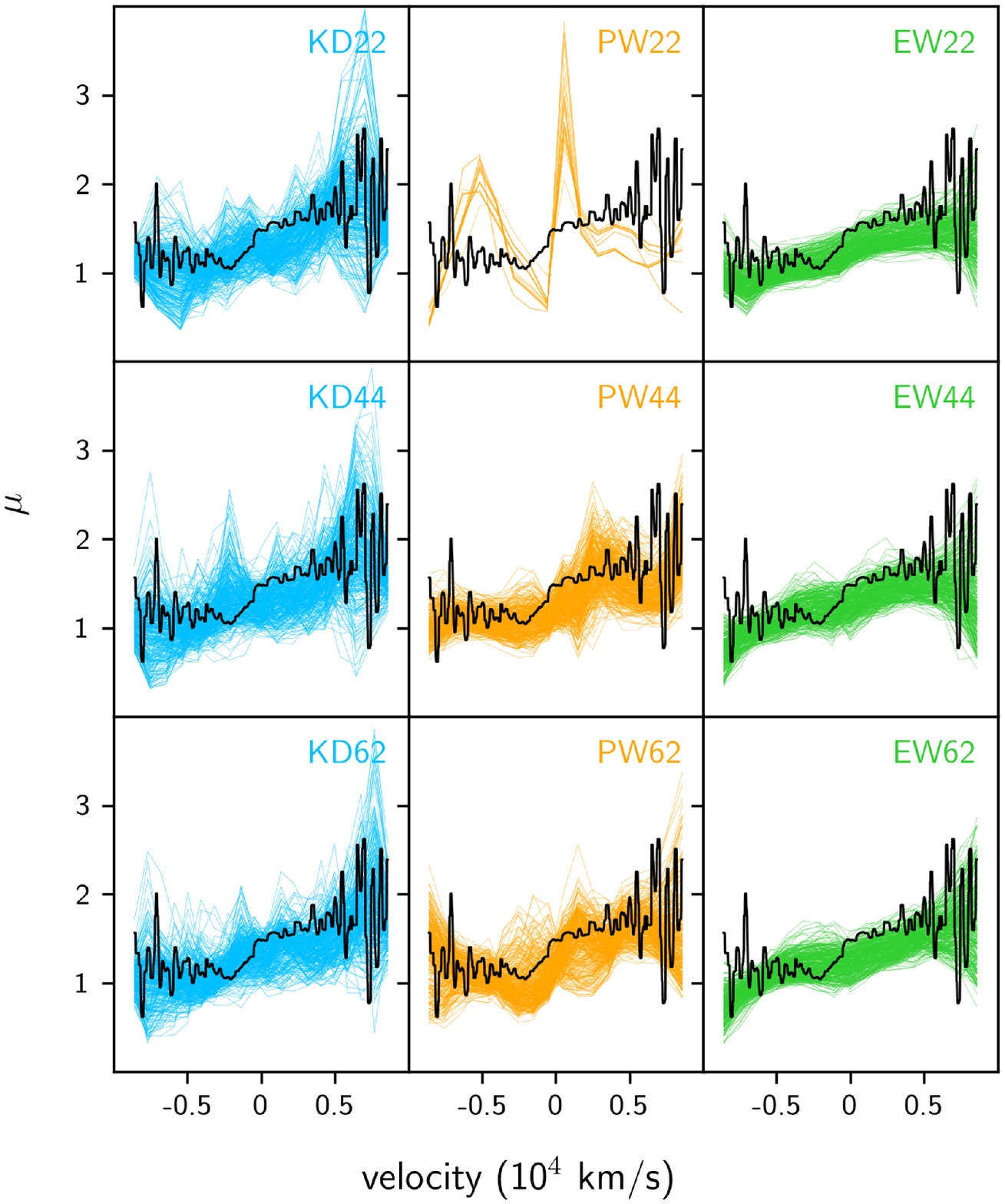}}
\caption{Randomly selected simulated magnification profiles $\mu(v)$ computed for the BLR models KD, PW, and EW, with an emissivity index $q=3$, inclinations of 22\degr, 44\degr,  and 62\degr, and microlensed by the magnification map rotated at 45\degr. The left panel illustrates $\mu(v)$ profiles from models only verifying the observational constraints on $\mu^{cont}$ and $\mu^{BLR}$, while the right panel illustrates $\mu(v)$ profiles also verifying the observational constraints on $WCI$ and $RBI$. The black line represents the observed $\mu(v)$ profile (see Fig.~\ref{fig:muwave}).}
\label{fig:he0435_muv}
\end{figure*}

In Fig.~\ref{fig:he0435_muv}, we show examples of simulated magnification profiles $\mu(v)$ constrained by the measured values of the magnification and distortion indices. As we can see, simulated profiles that verify the observational constraints on $WCI$ and $RBI$ reproduce the observed $\mu(v)$ profile reasonably well, especially if we remember that, in our simulations, we did not try to accurately reproduce the line profile itself. In particular, as discussed in Paper~I, the lack of flux in the core of the emission line coming from the biconical outflow seen at low inclination (PW22) can generate a small number of unrealistic simulations.

It is also interesting to note that most often microlensing only significantly affects large velocity chunks of the line profile. This is expected because a given caustic magnifies a slice of the BLR that is characterized by a range of projected velocities.

\end{appendix}

\end{document}